\input{aipcheck.tex}
\documentclass[final]{aipproc}
\layoutstyle{8x11single}
\def\be{\begin{equation}}
\def\ee{\end{equation}}
\def\bea{\begin{eqnarray}}
\def\eea{\end{eqnarray}}
\def\ba{\begin{array}}
\def\ea{\end{array}}

\begin{document}
\title[]{Magnetic moments of spin $\frac{1}{2}^+$ and spin $\frac{3}{2}^+$ charmed baryons}
\author{Harleen Dahiya}{address={Department of Physics, Dr. B.R. Ambedkar National Institute of Technology,
Jalandhar, Punjab-144 011, India.}}
\author{Neetika Sharma}{address={Department of Physics, Dr. B.R. Ambedkar National Institute of Technology,
Jalandhar, Punjab-144 011, India.}}
\author{P.K. Chatley}{address={Department of Physics, Dr. B.R. Ambedkar National Institute of Technology,
Jalandhar, Punjab-144 011, India.}}

\date{\today}

\begin{abstract}

The magnetic moments of spin $\frac{1}{2}{^+}$ and spin
$\frac{3}{2}{^+}$ charmed baryons  have been calculated in chiral
constituent quark model ($\chi$CQM). The effects of configuration
mixing and quark masses have also been investigated. The results
are not only in good agreement with existing experimental data but
also show improvement over other phenomenological models.

\end{abstract}

\maketitle
\section{Introduction}

Heavy flavor baryons play
an important role to understand the dynamics of light quarks in
the bound state as well as to understand QCD at the hadronic scale
\cite{garcilazo}. The phenomenological implications of the heavy quark component in 
the nucleon have been investigated to estimate the possible size 
of intrinsic charm (IC) content of the nucleon \cite{ic} as well as 
to calculate the static properties like masses, magnetic moment etc.  \cite{pdg} which
give valuable information regarding the internal structure of
baryons. 

The magnetic moments of spin $\frac{1}{2}{^+}$, spin
$\frac{3}{2}{^+}$ charmed baryons have been considered in
different approaches in literature. Calculations have been done
in the non-relativistic quark model \cite{choudhury,ccm}, 
Skyrme model \cite{skyrme}, bound state approach \cite{bsa}, 
relativistic three-quark model \cite{rtqm} etc.. More recently, magnetic moments 
have been studied
by considering the effective mass of the quark bound inside the baryon
\cite{patel}. Calculations 
for the charmed baryon magnetic moments have also been done in 
QCD sum rule method (QCDSR) \cite{wanglee}, 
QCD Spectral sum rule method (QSSR) \cite{qssr} and 
light cone QCD sum rule method (LCQSR) \cite{lcqsr,tam,tam3/2}.
However, there is little consensus among the different
model predictions of the magnetic moments of charmed baryons.

The {\it intrinsic} heavy quarks are created from the quantum fluctuations associated
with the bound state hadron dynamics and the process is completely
determined by nonperturbative mechanisms \cite{charm}.
It has been shown that one of the important model which finds application in
the nonperurbative regime is the chiral constituent quark model
($\chi$CQM) \cite{wein,eichten,cheng}. The $\chi$CQM with spin-spin generated
configuration mixing is able to give the satisfactory explanation
for the spin and flavor distribution functions \cite{hd,song}, hyperon $\beta$
decay parameters \cite{cheng}, strangeness content of the nucleon \cite{hds},
weak vector and axial-vector form factors \cite{ns}, octet and
decuplet baryon magnetic moments \cite{cheng1,johan,hdorbit} etc.. 
The successes of $\chi$CQM strongly suggest that constituent quarks
and the weakly interacting Goldstone bosons (GBs) provide the
appropriate degrees of freedom in the nonperturbative regime of
QCD. Thus, the quantum fluctuations generated by broken chiral
symmetry in $\chi$CQM should be able to provide a viable estimate
of the heavier quark flavor, in particular the $c \bar c$ \cite{charm,chengspin}. 

The purpose of the present paper is to estimate the
magnetic moments of spin $\frac{1}{2}{^+}$, spin $\frac{3}{2}{^+}$ 
charmed baryons in the SU(4) framework of $\chi$CQM. The generalized Cheng-Li
mechanism \cite{cheng1} has been incorporated to calculate explicitly the
contribution coming from the valence spin polarization, ``quark sea''
 polarization and its orbital angular momentum. 
Further, it would also be interesting to examine the effects of
the configuration mixing, symmetry breaking parameters,
confinement effects, quark masses etc. on the magnetic moments.

\section{Spin structure in chiral constituent quark model}
\label{cccm}

In this section, we briefly review the essentials of the
$\chi$CQM to calculate the spin structure of the baryons \cite{cheng1,johan,hdorbit}.
The basic process in the $\chi$CQM \cite{wein}
 is the internal emission of a Goldstone Boson
by a constituent quark which further splits into a $q \bar q$
pair as $q_{\pm} \rightarrow {\rm GB}^{0}+q'_{\mp} \rightarrow (q
\bar q^{'})+q'_{\mp}\,,$ where $q \bar q^{'} +q^{'}$ constitutes the
``quark sea'' \cite{cheng,hd,song,johan}. 
The effective Lagrangian describing interaction between quarks and
GBs is ${{\cal L}}= g_{15}{\bf \bar q}\left(\Phi\right) {\bf q}$, where
$g_{15}$ is the coupling constant, $I$ is the $4\times 4$ identity
matrix. The GB field $\Phi$ is expressed as {\footnotesize{
\bea \Phi = \left( \ba{ccccc} \frac{\pi^0}{\sqrt 2}
+\beta\frac{\eta}{\sqrt 6}+\zeta\frac{\eta^{'}}{4\sqrt
3}-\gamma\frac{\eta_c}{4} & \pi{^+} & \alpha K{^+} & \gamma
\bar{D}^0\\ \pi^- & -\frac{\pi^0}{\sqrt 2} +\beta
\frac{\eta}{\sqrt 6} +\zeta\frac{\eta^{'}}{4\sqrt 3}
-\gamma\frac{\eta_c}{4} &  \alpha K^0 & \gamma D^-\\ \alpha K^- &
\alpha \bar{K}^0  &  -\beta \frac{2\eta}{\sqrt 6}
+\zeta\frac{\eta^{'}}{4\sqrt 3} -\gamma\frac{\eta_c}{4} & \gamma
D^-_s\\ \gamma D^0  &\gamma D{^+}& \gamma D{^+}_s&
-\zeta\frac{3\eta^{'}}{4\sqrt 3} +\gamma\frac{3\eta_c}{4} \ea
\right).  \eea}} SU(4) symmetry breaking is introduced by
considering $M_c>M_s>M_{u,d}$ as well as by considering the masses
of GBs to be nondegenerate
$(M_{\eta_{c}}>M_{\eta^{'}}>M_{K,\eta}>M_{\pi})$. The parameter
$a(=|g_{15}|^2)$ denotes the transition probability of chiral
fluctuation of the splitting  $u(d) \rightarrow d(u)+ \pi^{+(-)}$,
whereas $a \alpha^2$, $a \beta^2$, $a \zeta^2$ and $a \gamma^2$
denote the probabilities of transitions of $u(d) \rightarrow s +
K^{-(o)}$, $u(d,s) \rightarrow u(d,s) + \eta$, $u(d,s) \rightarrow
u(d,s) + \eta^{'}$ and  $u(d) \rightarrow c +\bar{D}^0(D^-)$
respectively.

The spin structure of the baryon is defined as $\widehat B \equiv
\langle B|{\cal N}|B \rangle\,,$ where $|B\rangle$ is
the baryon wave function and ${\cal N}$ is the number operator defined as
$ {\cal N}=n_{u_{+}}u_{+} + n_{u_{-}}u_{-} + n_{d_{+}}d_{+} +
n_{d_{-}}d_{-} + n_{s_{+}}s_{+} + n_{s_{-}}s_{-} + n_{c_{+}}c_{+}
+ n_{c_{-}}c_{-}\,, $  $n_{q_{\pm}}$ being
the number of $q_{\pm}$ quarks \cite{cheng,hd,johan}. The ``quark sea'' contribution
to the total quark spin polarization ($\Delta q= q_{+}- q_{-}$) 
can be calculated by substituting for each valence quark
$ q_{\pm}\rightarrow \sum P_q q_{\pm} + |\psi(q_{\pm})|^2\,, $ 
where $\sum P_q$ is the probability of emission of GBs
from a $q$ quark and $|\psi(q_{\pm})|^2$ is the probability
of transforming a $q_{\pm}$ quark \cite{hdcharm}.
Using the spin and flavor wave functions for a given baryon, one
can easily find the spin structure and the spin polarizations.

The total wave function for the three quark system made from any of the 
$u$, $d$, $s$ or $c$ quarks is given as $|{\rm SU(8)}\otimes{\rm O(3)} \rangle = \phi \chi \psi$, 
where $\phi$ is a flavor wave function, $\chi$ is a spin wave function and $\psi$ 
is a spatial wave function.  The $SU(8)$ multiplets are decomposed into $SU(4) \otimes SU(2)$ 
multiplets and the details of the definition of the wave functions, 
can be found in \cite{yaoubook}. The spin structure of a spin $\frac{1}{2}{^+}$ and spin
$\frac{3}{2}{^+}$ baryons are respectively given as \be \hat B \equiv \langle B
|{\cal N}|B\rangle= {\cos}^2 \phi {\langle 120,^220_M|{\cal
N}|120,^220_M\rangle}_B + {\sin}^2 \phi {\langle 168,^220_M|{\cal
N}|168,^220_M\rangle}_B\,, \label{spinst} \ee  \be\hat B^{*}
\equiv \langle B^{*}|{\cal N}|B^{*}\rangle ={\langle 120,
^420_S|{\cal N}|120, ^420_S\rangle}_{B^{*}} \,. \label{spinst3/2}
\ee

\section{Magnetic moment in $\chi$CQM}
\label{magmom}

The magnetic moment of a given baryon receives contributions from
the valence quarks, ``quark sea'' and orbital angular momentum of
the ``quark sea'' 
\cite{cheng,hds,cheng1,hdorbit} and is expressed as \be 
\mu(B)_{{\rm total}} = \mu(B)_{{\rm val}} + \mu(B)_{{\rm sea}} +
\mu (B)_{{\rm orbit}}\,, \ee where $\mu
(B)_{{\rm val}}$ and $\mu (B)_{{\rm sea}}$ represent the
contributions of the valence quarks and the ``quark sea'' to the
magnetic moments due to spin polarizations. The term $\mu (B)_{{\rm
orbit}}$ corresponds to the orbital angular momentum contribution
of the ``quark sea''. In terms of quarks magnetic moments and spin polarizations, the
valence, sea and orbital contributions can be written as \be
\mu(B)_{{\rm val}}=\sum_{q=u,d,s,c} {\Delta q_{{\rm
val}}\mu_q}\,,~~ \mu(B)_{{\rm sea}}=\sum_{q=u,d,s,c} {\Delta
q_{{\rm sea}}\mu_q}\,,~~ \mu(B)_{{\rm orbit}}=\sum_{q=u,d,s,c}
{\Delta q_{{\rm val}}~\mu (q_{+} \rightarrow {q}_{-}^{'})}
\,,\label{mag} \ee where $\mu_q= \frac{e_q}{2 M_q}$ ($q=u,d,s,
c$) is the quark magnetic moment, $\mu (q_{+} \rightarrow
{q}_{-}^{'})$ is the orbital moment for any chiral fluctuation,
$e_q$ and $M_q$ are the electric charge and the mass respectively
for the quark $q$.

The valence and quark sea spin polarizations
($\Delta q_{\rm val}$ and $\Delta q_{\rm sea}$) can be
calculated for the baryons using the spin structure discussed in the previous section.
The orbital angular momentum contribution of each chiral
fluctuation is given as \cite{cheng,hdorbit} 
\be \mu (q_{+} \rightarrow
{q}_{-}^{'}) =\frac{e_{q^{'}}}{2M_q} \langle l_q \rangle +
\frac{{e}_{q} - {e}_{q^{'}}}{2 {M}_{{\rm GB}}}\langle {l}_{{\rm GB}}
\rangle\,,\ee
  where $\langle l_q
\rangle = \frac{{M}_{{\rm GB}}}{M_q+{M}_{{\rm GB}}}$ and $\langle
l_{{\rm GB}} \rangle=\frac{M_q}{M_q+{M}_{{\rm GB}}}\label{lq}$.
The quantities ($l_q$, $l_{{\rm GB}}$) and ($M_q$, ${M}_{{\rm
GB}}$) are the orbital angular momenta and masses of quark and
GBs, respectively. The orbital moment of each process 
is then multiplied by the probability for such a
process to take place to yield the magnetic moment due to all the
transitions starting with a given valence quark \be
[\mu (u_{\pm} \rightarrow )]=\pm a \left [ \left(\frac{1}{2}
+\frac{\beta^2}{6} + \frac{\zeta^2}{48}+ \frac{\gamma^2}{16}
\right) \mu (u_{+} \rightarrow u_-) +\mu (u_{+}\rightarrow d_-)
+\alpha^2 \mu (u_+ \rightarrow s_-) +\gamma^2 \mu (u_+ \rightarrow
c_-)\right ] \,, \label{muu} \ee \be [\mu (d_{\pm} \rightarrow )]
= \pm a \left [ \mu (d_{+} \rightarrow u_-) +\left (\frac{1}{2} +
\frac{\beta^2}{6} + \frac{\zeta^2}{48} + \frac{\gamma^2}{16}
\right) \mu (d_{+} \rightarrow  d_-) +\alpha^2 \mu (d_+
\rightarrow s_-) +\gamma^2 \mu (d_+ \rightarrow c_-) \right ] \,,
\label{mud} \ee
\be
[\mu (s_{\pm} \rightarrow )]= \pm a \left [\alpha^2 \mu (s_{+}
\rightarrow u_-) + \alpha^2 \mu (s_+ \rightarrow d_-) + \left
(\frac{2}{3} \beta^2+ \frac{\zeta^2}{48} + \frac{\gamma^2}{16}
\right) \mu (s_{+} \rightarrow s_- ) +\gamma^2 \mu (s_+
\rightarrow c_-)\right ] \,, \label{mus} \ee and
\be
[\mu (c_{\pm} \rightarrow )]= \pm a \left [\gamma^2 \mu (c_{+}
\rightarrow u_-) + \gamma^2 \mu (c_{+} \rightarrow d_-)+ \gamma^2
\mu (c_{+} \rightarrow s_-) +\left( \frac{3}{16}\zeta^2+
\frac{9}{16}\gamma^2 \right) \mu (c_{+} \rightarrow c_- ) \right ]
\,. \label{muc}
 \ee 
The above equations can easily be generalized by
including the coupling breaking and mass breaking terms and can be expressed
in terms of the $\chi$CQM parameters
($a, \alpha, \beta, \zeta, \gamma$), quark masses
($M_u,M_d,M_s,M_c$) and GB masses
($M_{\pi},M_{k},M_{\eta},M_{\eta'},M_{D},M_{D_s},M_{\eta_c}$).

\section{Results and Discussion}
\label{results}

Using the following set of $\chi$CQM parameters $a=0.12$, $\alpha
\simeq \beta= 0.45$, $\zeta = -0.21$ and $\gamma=0.11$ as well as
the on mass shell values of quarks and GBs \cite{mpi1,isgur}, 
we have calculated the magnetic moments of spin $\frac{1}{2}^+$ and spin
$\frac{3}{2}^+$ baryons in Tables \ref{spin1/2num}
and \ref{spin3/2num} respectively.  In the tables we have also presented the
available experimental data, the results of NRQM \cite{choudhury} and the 
results of other model calculations. 
From Table \ref{spin1/2num},
we find that our results compare fairly well 
with the experimental data  available for the octet
baryons. It is interesting to observe that our results in the case of
$p$, $\Sigma^{+}$, $\Xi^{0}$ and
$\Lambda^{0}$ give a perfect fit when compared with the experimental values
\cite{pdg} whereas for all other octet baryons our predictions are
within 10\% of the observed values. Since there is no experimental
information available in case of charmed baryon magnetic moments,
we compare our results with the predictions of QCD sum rule method
(QCDSR)\cite{wanglee}, Light Cone QCD sum rule method (LCQSR)
\cite{lcqsr}, QCD Spectral sum rule method (QSSR) \cite{qssr}. Our
results are found to be consistent with these approaches as well
as with the other models existing in literature. The explicit results
for the valence, sea and orbital contributions to the baryons
magnetic moments have been presented. A cursory look at the results in the 
table reveals that the sea and orbital contributions to the magnetic moments are
significant. The orbital part
contributes with the same sign as valence quark distribution,
while the sea part contributes with the opposite sign. However, the
sea and orbital contributions cancel each other to a large extent.
The sum of residual sea quark contribution and valence quark
contribution give the magnetic moment of baryons. Numerically
speaking, the sea quark contribution and orbital contributions are
quite large in magnitude except for $\Omega_{c}^0$,
$\Lambda_{c}{^+}$, $\Xi_{c}^{+}$, $\Xi_{c}^{0}$ and
$\Omega_{cc}^{+}$. It is also interesting to examine the role of
configuration mixing in spin $\frac{1}{2}{^+}$ baryon magnetic
moments. A detailed analysis of the configuration mixing
parameter $\phi$ reveals that the results with mixing are
in better agreement with the experimental data where the data
is available.

In Table \ref{spin3/2num},  we have compared our results for the spin $\frac{3}{2}{^+}$ baryons  
with other model calculations as well as with the available experimental data. 
Presently, only three experimental results are available for the decuplet baryons magnetic moments. 
Our predicted value for $\mu_{\Delta^{++}}$ is well within the experimental range \cite{pdg}. 
Similarly, in the case of $\mu_{\Delta{^+}}$  and $\Omega^{-}$, our predicted values agree with the experimental 
value \cite{kotulla,diehl}. In case of charmed baryons, there is no experimental 
information available, therefore, we have  compared our results with the predictions of the 
QCD sum rule \cite{wanglee} and Light Cone QCD sum rule \cite{tam3/2}. In this case also, we have presented the results for the valence, 
sea and the orbital contributions separately and we find that our predictions are in 
agreement with their results. There is a small discrepancy in
the case of $\Sigma^{*0}$ magnetic moment,
which is due to the significant sea contribution. 
The ``quark sea'' and orbital contributions are quite large in magnitude for all the charmed brayons except in the case of $\Omega^{*-}$, 
$\Omega_{c}^{*0}$, $\Omega_{cc}^{*+}$ and $\Omega_{ccc}^{*++}$. 
The measurements of the magnetic moments of charmed baryons represent an experimental challenge 
and several groups  BTeV, SELEX Collaboration are contemplating the possibility of performing it in the near future which
would test the success of present scheme.

\section{Summary and Conclusion}
\label{conc}

We have calculated the magnetic moments of spin $\frac{1}{2}{^+}$ and $\frac{3}{2}{^+}$ baryons  
in the framework of SU(4) $\chi$CQM. Without taking any of the magnetic moment as input, 
a considerable good fit is achieved in the case of the octet and decuplet baryons 
where the experimental data is available. In the case of charmed baryons, our results are consistent with the other approaches 
existing in the literature. The success of $\chi$CQM with the Cheng-Li mechanism and configuration 
mixing in achieving a fit to the magnetic moments suggest that constituent quarks and weakly interacting 
Goldstone Bosons provide the appropriate degree of freedom in the nonperturbative regime of QCD.

\section{ACKNOWLEDGMENTS}
H.D. would like to thank Department of Science and Technology, Government of India for financial
support.

\begin{table}
\footnotesize{
\begin{tabular}{lcccccccc}\hline
Baryon &Data&NRQM &QCDSR \cite{wanglee}&LCQSR & Valence & Sea & Orbital & Total\\
&\cite{pdg}&& QSSR\cite{qssr}& \cite{lcqsr} &&&& \\\hline
$p$&2.79$\pm 0.00$&3&2.82$\pm$0.26&2.7 $\pm$0.5&2.90&$-$0.58&0.47 &2.80\\
$n$&$-$1.91$\pm$0.00&$-$2 &$-$1.97 $\pm$ 0.15&$-1.8\pm$0.35 & $-$1.85&0.18&$-$0.44&$-$2.11\\
$\Sigma^{+}$&2.458$\pm$0.010&2.88 &2.31 $\pm$0.25&2.2$\pm$0.4&2.51&$-$0.51&0.40&2.39\\
$\Sigma^{0}$&-- &0.88& 0.69 $\pm$0.07& 0.5$\pm$ 0.10&0.74&$-$0.22&0.02& 0.54\\
$\Sigma^{-}$&$-$1.160 $\pm$0.025& $-$1.12&$-$1.16 $\pm$0.10& $-0.8\pm 0.2$& $-$1.02&0.06&$-$0.36&$-$1.32\\
$\Xi^{0}$& $-$1.250 $\pm$0.014 &$-$ 1.53& $-$1.15$\pm$ 0.05&$-1.3 \pm$0.3 &$-$1.29&0.14&$-$0.09&$-$1.24\\
$\Xi^{-}$&$-$0.6507$\pm$0.0025& $-$0.53&$-$0.64$\pm$ 0.06& $-0.7 \pm 0.2$&$-$0.59&0.03&0.06& $-$0.50\\
CSGR&0.49$\pm$ 0.05 &0.0&&&&&&0.46\\
$\Lambda^{0}$&$- 0.613 \pm$0.004&$-$0.65 &$- 0.56\pm$0.15&$-0.7 \pm 0.2 $&$-$0.59&$-$0.06&$-$0.01& $-$0.66\\
 \hline
$\Sigma_{c}^{++}$&...& 2.54&2.1$\pm$ 0.3 &...&2.32&$-$0.52&0.40&2.20 \\
$\Sigma_{c}^{+}$&...& 0.54&0.6 $\pm$0.1 &...&0.51&$-$0.23&0.02&0.30\\
$\Sigma_{c}^{0}$&...& $-$1.46&$-1.6 \pm 0.2$&...& $-$1.30&0.06&$-$0.36&$-$1.60\\
$\Xi_{c}^{'+}$&...& 0.77&...&...& 0.77&$-$0.21&0.19&0.76 \\
$\Xi_{c}^{'0}$&...& $-$1.23& ...&...& $-$1.16&0.03&$-$0.19&$-$1.32\\
$\Omega_{c}^{0}$&...& $-$0.99&...&...&$-$0.93&0.04&$-$0.01&$-$0.90 \\
$\Lambda_{c}^{+}$&...& 0.39&0.15$\pm$ 0.05& 0.40$\pm$ 0.05 & 0.409&$-$0.019&0.002&0.392 \\
$\Xi_{c}^{+}$&...&0.39&...&0.50 $\pm$0.05 &0.41&$-$0.02& 0.01&0.40  \\
$\Xi_{c}^{0}$&...& 0.39&...&0.35$\pm$ 0.05&0.29&$-$0.0003&$-$0.01&0.28\\ \hline
$\Xi_{cc}^{++}$&...&$-$0.15&...&...&0.025&0.111&$-$0.080&0.006\\
$\Xi_{cc}^{+}$&...&0.85&...&...&0.79&$-$0.02&0.07&0.84\\
$\Omega_{cc}^{+}$&...& 0.73&...&...& 0.706&$-$0.012&$-$0.004&0.697\\
\hline
\end{tabular}\caption{ Magnetic moment of spin $\frac{1}{2}{^+}$ charmed baryons with configuration mixing (in units of $\mu_{N}$).} \label{spin1/2num}}
\end{table}

\begin{table}
\footnotesize{
\begin{tabular}{lcccccccc}\hline
Baryon  &Data&NRQM &QCDSR&LCQSR& Valence & Sea & Orbital & Total\\
&\cite{pdg}&& \cite{wanglee} & \cite{tam3/2}&&&& \\\hline
$\mu_{\Delta^{++}}$&3.7 $\sim$ 7.5& 6& 4.13$\pm$1.30&4.4$\pm$ 0.8&4.53&$-$0.97&0.95&4.51\\
$\mu_{\Delta^{+}}$& $2.7^{+1.0}_{-1.3} \pm 1.5\pm 3$ \cite{kotulla}& 3& 2.07$\pm$0.65& 2.2$\pm$0.4&2.27&$-$0.61&0.34&2.00\\
$\mu_{\Delta^{0}}$&...&0.0 &0.0&0.0& 0.0& $-$0.25&$-$0.26&$-$0.51\\
$\mu_{\Delta^{-}}$&...& $-$3&$- 2.07\pm$0.65&$-$2.2$\pm$0.4& $-$2.27& 0.12& $-$0.87&$-3.02$\\
$\mu_{\Sigma^{*+}}$&...&3.35&2.13$\pm$0.82 & 2.7$\pm$0.6&2.74&$-$0.67&0.62&2.69\\
$\mu_{\Sigma^{*0}}$&...&0.35&$0.32\pm0.15$&0.20$\pm$0.05& 0.29& $-$0.29& 0.02&0.02\\
$\mu_{\Sigma^{*-}}$&...& $-$2.65&$-1.66\pm0.73$ &$-$2.28$\pm$0.5&$-$2.16& 0.11& $-$0.59&$-$2.64\\
$\mu_{\Xi^{*0}}$&...&0.71&$-0.69\pm 0.29$&0.40$\pm$0.08& 0.51& $-$0.26& 0.29&0.54\\
$\mu_{\Xi^{*-}}$&...&$-$2.29 &$-1.51 \pm 0.52$ & $-$2.0$\pm$0.4&$-$1.64&0.08& $-$0.31&$-$1.87\\
$\mu_{\Omega^{*-}}$&$-$2.02 $\pm$0.06 &$-$1.94&$-1.49 \pm 0.45$&$-$1.65$\pm$0.35& $-$1.76& 0.08&$-$0.03& $-$1.71\\
& $-$1.94 $\pm$ 0.31 \cite{diehl}&& &&&&&\\\hline
$\mu_{\Sigma_{c}^{*++}}$&...& 4.39&....&$4.81\pm1.22$ &4.09&$-$0.80&0.63&3.92\\
$\mu_{\Sigma_{c}^{*+}}$&...& 1.39&...&$2.00\pm0.46$ &1.30&$-$0.36&0.03&0.97\\
$\mu_{\Sigma_{c}^{*0}}$&...&$-$1.61&...&$-0.81\pm0.20$& $-$1.50&0.09&$-$0.58&$-$1.99\\
$\mu_{\Xi_{c}^{*+}}$&...&1.74&...&$1.68\pm0.42$& 1.67&$-$0.39&0.31 &1.59\\
$\mu_{\Xi_{c}^{*0}}$&...&$-$1.26&...&$-0.68\pm0.18$& $-$1.21& 0.08& $-$0.30&$-$ 1.43\\
$\mu_{\Omega_{c}^{*0}}$&...& $-$0.91&...&$-0.62\pm0.18$& $-$0.89& 0.05&$-$0.02&$-$0.86\\\hline
$\mu_{\Xi_{cc}^{*++}}$&...&2.78&...&...&2.78&$-$0.44& 0.32&2.66\\
$\mu_{\Xi_{cc}^{*+}}$&...&$-$0.22&...&...&$-$ 0.22&0.04&$-$0.29&$-$0.47\\
$\mu_{\Omega_{cc}^{*+}}$&...& 0.13&...&...&0.13&0.02&$-$0.01&0.14\\ \hline
$\mu_{\Omega_{ccc}^{*++}}$&...& 1.17&...&...&0.165&0.011&$-$0.002&0.155\\
\hline
\end{tabular}\caption{The magnetic moments of the spin $\frac{3}{2}{^+}$ charmed baryons (in units of $\mu_{N}$).} \label{spin3/2num}}
\end{table}

\end{document}